# Heat Transport in Silicon Nitride Drum Resonators and its Influence on Thermal Fluctuation-induced Frequency Noise

Nikaya Snell, Chang Zhang, Gengyang Mu, Alexandre Bouchard, Raphael St-Gelais[†]
*Department of Mechanical Engineering, University of Ottawa, Ottawa, ON, Canada*

Silicon nitride (SiN) drumhead resonators offer a promising platform for thermal sensing due to their high mechanical quality factor and the high temperature sensitivity of their resonance frequency. As such, gaining an understanding of heat transport in SiN resonators as well as their sensing noise limitations is of interest, both of which are goals of the present work. We first present new experimental results on radiative heat transport in SiN membrane, which we use for benchmarking two recently proposed theoretical models. We measure the characteristic thermal response time of square SiN membranes with a thickness of 90 ± 1.7 nm and side lengths from 1.5 to 12 mm. A clear transition between radiation and conduction dominated heat transport is measured, in close correspondence with theory. In the second portion of this work, we use our experimentally validated heat transport model to provide a closed-form expression for thermal fluctuation-induced frequency noise in SiN membrane resonators. We find that, for large area SiN membranes, thermal fluctuations can be greater than thermomechanical contributions to frequency noise. For the specific case of thermal radiation sensing applications, we also derive the noise equivalent power resulting from thermal fluctuation-induced frequency noise, and we show in which conditions it reduces to the classical detectivity limit of thermal radiation sensors. Our work therefore provides a path towards achieving thermal radiation sensors operating at the never attained fundamental detectivity limit of bolometric sensing. We also identify questions that remain when attempting to push the limits of radiation sensing, in particular, the effect of thermal fluctuation noise in closed-loop frequency tracking schemes remains to be clarified.

## I. INTRODUCTION

A variety of sensing technologies have been built using nanomechanical resonators, such as mass [1]–[3], force [4], [5], and thermal radiation sensors [6]–[12]. The use of silicon nitride (SiN) resonators in thermal sensing schemes is especially appealing due to the ability to form high quality factor resonators whose resonance frequency is highly sensitive to temperature changes [7], [8], [13]–[15]. To further the development of SiN thermal sensors, there has been recent interest in understanding the radiative and conductive heat transport in SiN membranes [8], [15]–[17].

Recent studies [8], [15], including one by the authors of the current work [8], proposed models for thermal radiation heat transport in square silicon nitride membranes using either a circular approximation [8] or a first-order Taylor expansion approximation [15]. A distinctive feature of our work [8] was the inclusion an analytical model for the emissivity of SiN. On the other hand, experimental validation of our model was only preliminary. A single membrane size was tested experimentally using a static radiator, which resulted in relatively large uncertainties due to drift. In [15], a first order Taylor model corresponded well with experimental measurements for 50 nm thick square membranes. Dynamic measurements were performed for various membranes side-lengths up to 4 mm, for which the fraction of heat transfer occurring by radiation ($x_{rad}$) is predicted to be approximately 70% [8], i.e., where 30% of the total heat transfer still occurs via conduction. Despite this conclusive demonstration, observation of a fully radiation-dominated region is still desirable. This would notably allow a more direct confirmation of existing models [8], [18] and measurements [15] on the emissivity of SiN, without the need to account for conduction heat transfer.

The goal of this work is two-fold. We first provide additional experimental validation for these models by performing dynamic measurements of the thermal response time of SiN membranes for membrane sizes up to 12 mm. We observe a clear region of radiation-dominated heat transport ($x_{rad} \approx 90\%$) that allows for a detailed comparison with recently proposed models for thermal transport [8], [15], emissivity [8], [18], and thermal conductivity [16], [17], [19].

Secondly, we use our validated model for heat transport to derive a closed-form expression for thermal fluctuation-induced frequency noise in SiN resonators. We show that thermal fluctuation noise can dominate over thermomechanical frequency noise in large area SiN. This outlines the importance of considering thermal fluctuation noise together with other noise sources (i.e., thermomechanical and detection noise) within recently proposed frameworks for frequency noise in nanomechanical resonators [20]–[24].

---

[†] raphael.stgelais@uottawa.ca

## II. EXPERIMENTAL METHODS

The SiN drumhead resonators used in this research are fabricated using commercially available silicon wafers with ~100 nm thick low pressure chemical vapour deposition (LPCVD) low stress SiN. The wafers are patterned using photolithography and individual chips, with square membrane side-length sizes ($L$) between 1 mm – 12 mm, are then released by KOH etching. Post-fabrication, the thickness ($d$) of 12 different SiN films are measured using an ellipsometer (Horiba Ltd. UVISEL FUV-NIR), yielding thicknesses of 90 ± 1.7 nm. The actual membrane side lengths are then quantified by correcting for KOH over-etching. The over-etching magnitude is assessed, for every chip, by measuring relative change of the membrane frame dimension using a micrometer (Mitutoyo Inc.). A membrane size error of approximately 15 μm is estimated from the repeatability of this measurement.

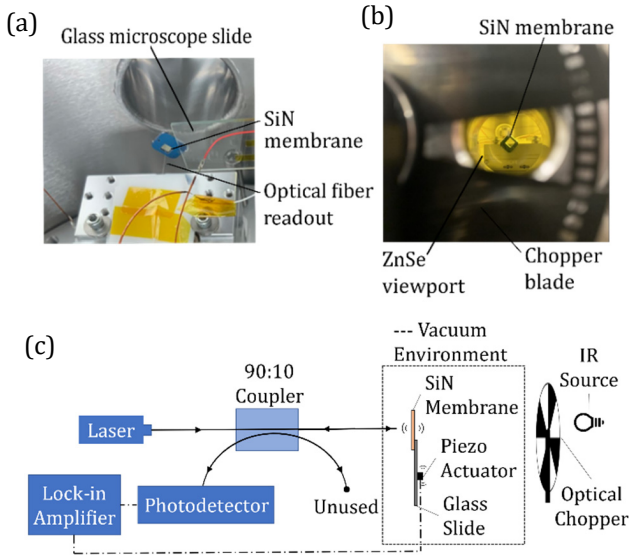

FIG. 1. (a) SiN resonator mounted on a glass slide with a piezo actuator (not pictured). (b) View of SiN resonator from outside the vacuum chamber through the optical chopper and the ZnSe viewport. (c) Schematic of interferometer, resonator mounting, and IR source modulation.

Experiments are performed in a custom-made high vacuum testing chamber (~7.5 × $10^{-7}$ Torr). The corner of the resonator's silicon frame is mounted on a glass microscope slide using Crystalbond™ shown in FIG. 1 (a). Also attached to the glass slide is a shear piezo used to actuate the membrane. A fiber optic interferometer [25] records the resonator vibration amplitude and shifts in the resonance frequency. The interferometer consists of a 1550 nm Orion™ laser, a 90:10 optical fiber coupler, and an amplified photodetector (Thorlabs Inc. PDA20CS2) as seen in the schematic in FIG. 1 (c). Signal from the photodetector is sampled using a Zurich Instrument Ltd. MFLI Lock-in amplifier. The lock-in amplifier is also used to actuate the shear piezo in a phase locked loop (PLL) that drives the membrane at its resonance frequency. Thermal infrared excitation of the membrane is performed with an infrared light source (ArcOptix S.A. ArcLight 1 – 25 μm IR source) located outside the vacuum chamber. The vacuum chamber is equipped with a zinc selenide (ZnSe) viewport from Thorlabs Inc. to allow transmission in the 600 nm - 16 μm spectral range (FIG. 1b). Modulation of the IR source is performed using a Thorlabs Inc. optical chopper equipped with a 10% duty cycle blade.

## III. HEAT TRANSPORT IN SIN DRUM RESONATORS

Two theoretical models were recently proposed for predicting thermal transport in drumhead resonators; both of these models can be applied to square drumhead resonators and include heat transfer by conduction and radiation [8], [15]. The model proposed in [8] uses a circular approximation with an effective radius ($r_{eff} = 1.252\, L/2$) to solve the heat equation in the square membrane. This model yields a characteristic thermal response time:

$$\tau_{th} = \frac{c_p \rho d}{8\sigma_{SB}\varepsilon T^3} x_{rad}, \quad (1)$$

where $c_p = 700\, \frac{J}{kgK}$ [26] is the specific heat capacity, $\rho = 2900\, kg/m^3$ [27] is the material density, $d$ is the membrane thickness, $\sigma_{SB}$ is the Stefan-Boltzmann constant, $\varepsilon$ is the total hemispherical emissivity of the membrane, and $T$ is the ambient temperature. Also necessary for calculating the thermal time constant is the fraction ($x_{rad} = G_{rad}/G$) of the membrane radiative conductance ($G_{rad}$) over the total conductance ($G$, in units W/K). For a square membrane this is calculated as [8]

$$x_{rad} = 1 - \frac{2}{\beta r_{eff}} \frac{I_1(\beta r_{eff})}{I_0(\beta r_{eff})}, \quad (2)$$

where $I_N$ is the Nth-order modified Bessel function of the first kind, and $\beta$ is given by

$$\beta = \sqrt{\frac{8\sigma_{SB}\varepsilon T^3}{kd}}, \quad (3)$$

where $k$ is the material conductivity of the membrane. In turn, the total conductance ($G$) between the membrane and its environment can be evaluated from $x_{rad}$ and $G_{rad}$:

$$G = \frac{4\sigma_{SB}\varepsilon T^3 A}{x_{rad}}, \quad (4)$$

where $A = 2L^2$ for a suspended membrane exposed from both sides.

Conversely, the thermal response time model found in [15] solves for the square membrane temperature profile, and truncates the solution to the first-order Taylor expansion term:

$$\tau_{th} = \left(\frac{2\pi^2}{L^2}\frac{\sum_i d_i k_i}{\sum_i d_i \rho_i c_{p_i}} + \frac{8\sigma_{SB}\varepsilon T^3}{\sum_i d_i \rho_i c_{p_i}}\right)^{-1}, \quad (5)$$

where $i$ indicates the layer number in the case of a multilayer assembly.

These models are compared to experiments by exposing SiN membranes to IR radiation, which causes their temperature to increase, resulting in a measurable resonance frequency shift $\alpha$, in Hz/K, given by [8]:

$$\alpha \approx -\frac{E\alpha_T}{2\sigma(1-\nu)}f_r \quad (6)$$

where $\alpha_T$ is the material coefficient of thermal expansion ($\sim 2.2 \times 10^{-6}$ K$^{-1}$ for SiN), $E$ is Young's modulus, $\sigma \approx 100$ MPa is the membrane's built-in tensile stress at room temperature, $\nu$ is the Poisson ratio, and $f_r$ is the resonance frequency of the excited membrane eigenmode. Dividing this expression by $G$ yields the responsivity to absorbed radiation, in Hz/W:

$$\Re_{abs} = \frac{\alpha}{G}H_{th}(j\omega), \quad (7)$$

where $H_{th}(j\omega)$ is a one pole low-pass filter that accounts for the thermal response time of the resonator:

$$H_{th}(j\omega) = \frac{1}{1 + j\omega\tau_{th}} \quad (8)$$

The responsivity to the incident radiation intensity, $\Re$, is finally given by multiplying $\Re_{abs}$ by the membrane absorption coefficient, which, according to Kirchhoff's law, equals membrane emissivity:

$$\Re = \varepsilon_{\lambda,\theta}(\lambda, 0)\Re_{abs}, \quad (9)$$

where $\varepsilon_{\lambda,\theta}(\lambda, 0)$ is the spectral normal emissivity. As shown in [8] and [18], for the case of a plain freestanding SiN membrane, the spectral normal emissivity $\varepsilon_{\lambda,\theta}(\lambda, 0)$ and the angle-integrated spectral hemispherical emissivity $\varepsilon_\lambda(\lambda)$ can be approximated as equal.

In order to benchmark models given in [8], [15], we characterize 90 ± 1.7 nm thick SiN membranes with the following nominal side-lengths ($L$): 1.5 mm, 3 mm, 6 mm, and 12 mm. Once the internal oscillator of the lock-in is phase-locked to one of the membrane's eigenmodes, the membrane is intermittently exposed to our IR source at a rate of 4 Hz using the chopper system that blocks light for 90% of the cycle.

In FIG. 2 (a), we clearly resolve a resonance frequency shift of ~9 Hz for resonance mode (3, 4), which is centered at around 76 kHz, for a $L = 6$ mm membrane. This would correspond to ~1.4 μW of absorbed radiation according to Eq. (7), in cases where $\omega \ll \frac{1}{\tau_{th}}$. In FIG. 2 (b), we measure a 0.006 Hz RMS residual frequency shift for mode (3, 4). This residual shift corresponds to a ~0.9 nW RMS sensor absorbed noise floor for a sensing bandwidth of 50 Hz, which is the lowest value achieved in this work.

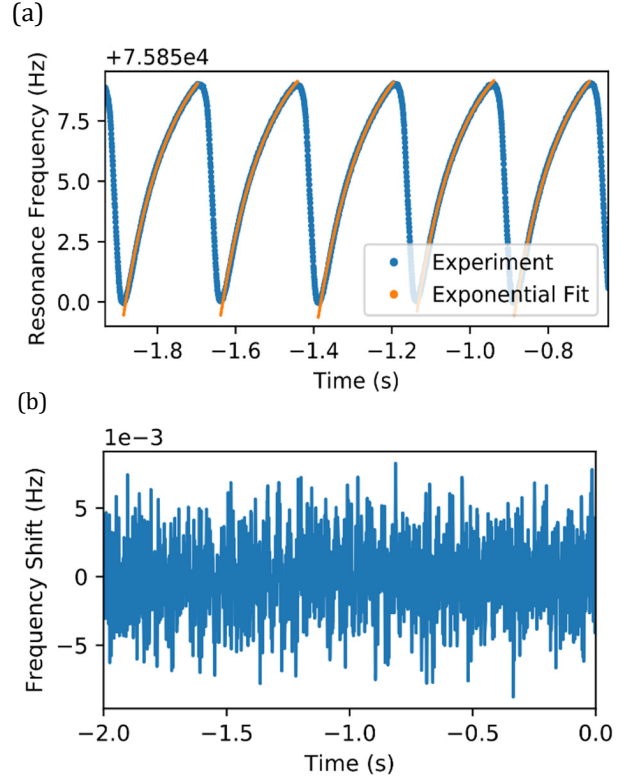

FIG. 2. (a) Frequency response of mode (3, 4) of an $L$ = 6 mm membrane as it is intermittently exposed to an IR source. Upon exposition to the IR source, the SiN stress reduces, thus lowering resonance frequency. (b) 0.006 Hz RMS noise measured when the membrane is not exposed to radiation.

Changes in resonance frequency over time are fit with exponential curves, as seen in FIG. 2 (a), to extract the characteristic thermal response time, $\tau_{th}$, for multiple membrane dimensions. Accurate measurements of $\tau_{th}$ are ensured by configuring the PLL settings to a high bandwidth, such that the observed response time is limited solely by thermally-induced changes in the membrane resonance frequency. To determine these settings, the thermal time constant of various membranes were measured with multiple PLL bandwidths. As can be seen in FIG. 3 for the case of an $L = 6$ mm membrane, there are discrepancies in the value of the thermal time constant of the resonator at low PLL bandwidths. PLL bandwidths lower than 20 Hz were too slow to accurately measure the resonator's thermal time constant, whereas a plateau region can be seen for larger bandwidths. Considering this effect, the PLL bandwidth for all measurements is set to ≥50 Hz, which was verified to be fast enough for all membrane sizes and eigenmodes used in this work. The demodulator bandwidth of the lock-in amplifier is set to five times the PLL bandwidth, i.e., to 250 Hz.

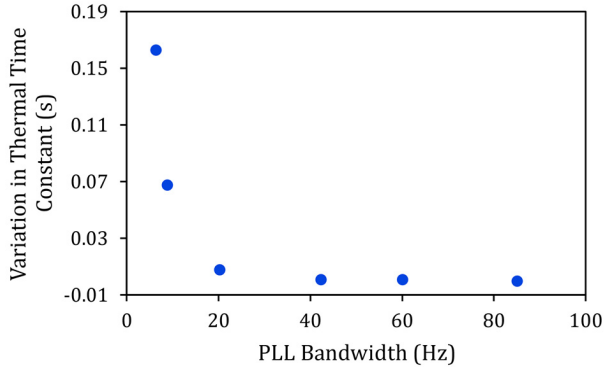

FIG. 3. PLL calibration. Increasing the PLL bandwidth until the measured response time shows a plateau, indicating that the PLL bandwidth is sufficiently large for measuring thermal response time of the membrane. Measurements performed for mode (2, 2) of a SiN membrane of 6 mm side-length.

Thermal time constants measured for multiple membrane sizes are shown in FIG. 4 (a) and show relatively good correspondence with both theoretical models [8], [15]. The model curves are generated assuming hemispherical emissivity ($\varepsilon$) of the 90 ± 1.7 nm thick SiN membrane is 0.097 in accordance with the emissivity model proposed in [8] and later corrected in [18]. Furthermore, different conductivities ($k$) are used when calculating the models to illustrate the effect of the large variability of values reported in literature [16], [17], [19] for this parameter. The value of $k$ reported in the most recent comparative study (2.7 W/mK [17]) yields the best correspondence with both heat transfer models, which further validates these recent conclusions [17]. We therefore assume $k = 2.7$ W/mK for the remainder of this work. In FIG. 4 (b), we present the experimental data for additional eigenmodes and we note that higher mode orders have slightly lower thermal time constant values when compared to mode (1, 1). This can be attributed to the mode shape; wherein higher order modes have antinodes that are closer to the heat-dissipating Si frame.

In FIG. 4 (b), both models [8], [15] predict relatively accurate thermal time constants for small membrane sizes ($L$); however, we note a systematic difference between the two existing models in the radiation dominated regions ($L \geq 6$ mm). It appears that keeping only the first-order Taylor term for the membrane temperature profile, as in [15], can lead to an overestimation of $\tau_{th}$ as it cannot properly model a fully radiation-dominated profile that is expected to have a box-like shape—i.e., constant temperature almost everywhere, and sharp changes in temperature near the membrane edges. We note however that the over estimation of is modest (~20%) and, in most practical case, it is likely to be within the error margin resulting from other factors.

We emphasize that no fit parameters are used in FIG. 4 (b) for matching measurement with theory curves. This indicates that the emissivity model for SiN proposed in [8] and [18], using the permittivity model given in [28], can likely be used with a high confidence level.

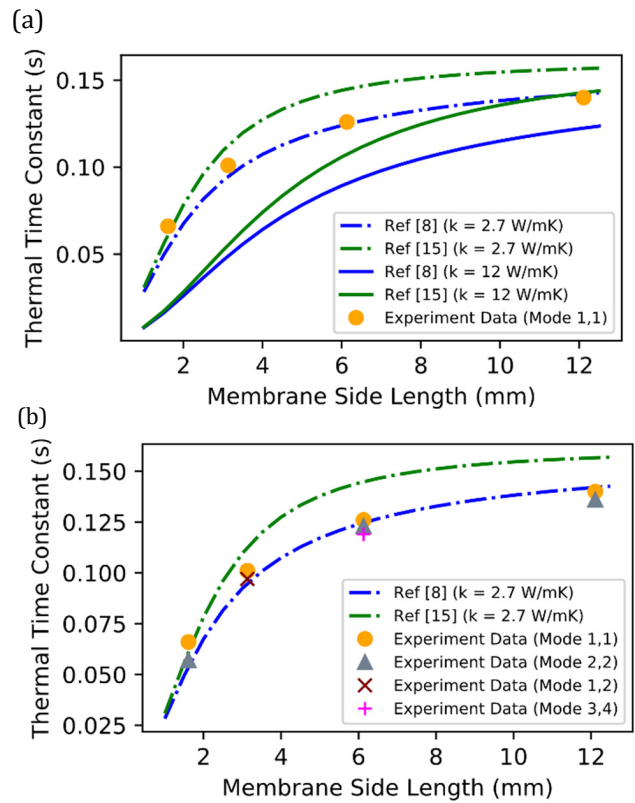

FIG. 4. Characteristic thermal response time of SiN resonators. (a) Comparison of experimental thermal time constant data to the Circular approximation model [8] and first order Taylor expansion approximation [15] for various membranes sizes. The thermal conductivity (k) of SiN is varied to show the impact different reported values. (b) Additional vibration mode measurements compared with thermal time constant models from [8], [15] assuming the most recent literature value for thermal conductivity ($k = 2.7$ W/mK [17]) and $\varepsilon = 0.097$. A radiation dominant region is obvious for membrane side lengths ≳6 mm. For readability, only averaged measurements for multiple experimental trials are plotted in both figures. Error bars have been included to account for measurement

repeatability, but the standard deviation values (on the order of 0.01 mm [x-axis] and 0.001 s [y-axis]) are too small to be visible.

## IV. THERMAL FLUCTUATION-INDUCED FREQUENCY NOISE

Armed with our experimentally validated thermal transport model, we can now derive expressions for frequency noise and for noise equivalent power ($NEP$) resulting from thermal fluctuation noise in frequency-shift based SiN radiation sensors. We also analyse the contribution of thermal fluctuation noise relative to others—namely thermomechanical noise and experimental detection noise—that have been more extensively discussed in previous work [20]–[22], [29].

Thermal fluctuation noise results from random energy exchanges between the resonator and its surrounding environment and therefore depends on the thermal conductance ($G$) of the structure. These exchanges affect the resonator temperature, which in turn results in resonance frequency shifts. We can express the fractional frequency ($y = \frac{\Delta f}{f}$) fluctuation power density spectrum due thermal noise by multiplying the temperature noise spectrum ($S_T$, in K$^2$/Hz) [6], [23] by the normalized temperature sensitivity $\alpha^2/f_r^2$:

$$S_{y,th}(\omega) = S_T(\omega)\frac{\alpha^2}{f_r^2} = \frac{4k_BT^2}{G}|H_{th}(j\omega)|^2\frac{\alpha^2}{f_r^2}, \quad (10)$$

where $k_B$ is the Boltzmann constant. In turn, by combining this noise density function with the sensor responsivity [Eq. (9)], we can express the sensor noise equivalent power due to thermal fluctuation as:

$$NEP_{th} = \frac{\sqrt{S_{y,th}}f_r}{|\Re|} = \frac{\sqrt{4k_BT^2G}}{\varepsilon_{\lambda,\theta}(\lambda,0)}, \quad (11)$$

Eq. (11), combined with the expression of $G$ given in Eq. (4), provides a closed-form expression for the fundamental $NEP$ limit of drumhead resonators when used as thermal radiation sensors. Note that when calculating $G$ for use in Eq. (11), the area $A$ must account for both sides of the membrane (i.e., $A = 2L^2$) unless surface functionalization allows for radiative coupling solely on the front surface. Care must also be given not to confuse spectral normal emissivity $\varepsilon_{\lambda,\theta}(\lambda,0)$ with the total hemispherical emissivity $\varepsilon$, especially in cases, such as for SiN, where the former is strongly wavelength dependant [8], [18].

We note that Eq. (11) reduces to the classical fundamental detectivity limit of bolometers in the idealized case of a front-side coupled membrane ($A_{tot} = A_{front} = L^2$) with perfect absorption ($\varepsilon_{\lambda,\theta} = \varepsilon = 1$) and $x_{rad} = 1$ [30]:

$$D_{th}^* = \frac{\sqrt{A_{front}}}{NEP_{th}} = (16k_BT^5\sigma_{SB})^{-\frac{1}{2}}, \quad (12)$$

which is approximately $1.8 \times 10^{10} \frac{cm\sqrt{Hz}}{W}$ at $T = 300\ K$.

Obviously, other sources of noise must be accounted for in practical applications and thermomechanical noise is often the dominant one as indicated in [20]–[22], [29]. The spectral density of frequency noise induced by thermomechanical fluctuations is given by [20], [21]:

$$S_{y,tmech}(\omega) = \frac{k_BT}{m_{eff}\omega_r^3Qa_{rss}^2}|H_{tmech}(j\omega)|^2, \quad (13)$$

where $Q$ is the quality factor of the resonator, $m_{eff}$ is the effective resonator mass, $\omega_r$ is the angular resonance frequency, and $a_{rss}$ is the amplitude of the resonator motion in steady state. $H_{tmech}(j\omega)$ denotes the resonator transfer function, which is a one-pole low pass filter similar to Eq. (8) but with a characteristic resonator time constant $\tau_{mech} = Q/(\pi f_r)$ instead of the thermal time constant ($\tau_{th}$). Minimization of thermomechanical noise is usually achieved by using a driving amplitude at the onset of nonlinearity [31]:

$$a_{rss} = a_{crit} = 0.56\frac{L}{\sqrt{Q}}\sqrt{\frac{\sigma}{E}}. \quad (14)$$

Using the same approach as for $NEP_{th}$, we can derive an expression for thermomechanical $NEP$ using $S_{y,tmech}$ in lieu of $S_{y,th}$ in Eq. (11), and the expression for $a_{rss}$ given in Eq. (14):

$$NEP_{tmech} = \frac{3.57\ G(1-\nu)}{\varepsilon_{\lambda,\theta}(\lambda,0)\ \alpha_{T}L}\sqrt{\frac{k_BT\sigma}{m_{eff}\omega_r^3E}}\frac{|H_{tmech}(j\omega)|}{|H_{th}(j\omega)|}. \quad (15)$$

Another noise source that affects sensing performance is detection noise, which is specific to each experimental apparatus. A convenient way of expressing detection noise is to quantify its magnitude relative to thermomechanical noise [20], [22]:

$$S_{y,det}(\omega) = \kappa_d^2\frac{k_BT}{m_{eff}\omega_r^3Qa_{rss}^2}, \quad (16)$$

where $\kappa_d$ is a dimensionless scaling parameter that indicates whether the background detection noise is sufficiently low for resolving thermomechanical displacement fluctuations (i.e., if $\kappa_d < 1$). Measurements of $\kappa_d$ are presented in supplementary section S1.

It is insightful to evaluate the relative contribution of $NEP_{th}$ relative $NEP_{tmech}$, which commonly dominates frequency noise in nanomechanical resonators [20]–[22],

[29]. We explore the relative contribution of $NEP_{th}$ by defining the dimensionless ratio $\eta$:

$$\eta = \frac{NEP_{th}(\omega)}{NEP_{tmech}(\omega)} = \left(\frac{S_{y,th}}{S_{y,tmech}}\right)^{1/2}$$
$$= \frac{0.56\alpha_t L}{(1-v)}\sqrt{\frac{m_{eff}\omega_r^3 TE}{G\sigma}}\frac{|H_{th}(j\omega)|}{|H_{tmech}(j\omega)|} \ . \quad (17)$$

By evaluating $\eta$ (see FIG. 5 a) for parameters of a typical large area membrane ($L = 6$ mm) used in this work, we surprisingly find that thermal fluctuation noise can dominate over thermomechanical noise (i.e., $\eta > 1$), especially in large area SiN membranes. In FIG. 5 (b), we note that thermal fluctuation noise can even dominate for membrane sizes down to 100 µm if we consider driving the commonly employed mode (2, 2) at the onset of non-linearity, with a typical Q-factor of 1 million. Frequency noise dominated by thermal fluctuation noise is in striking contrast with recent work on frequency noise in SiN string resonators, which clearly demonstrated thermomechanical fluctuation-dominated behaviour [22]. Our current findings do not contradict this recent study, for which the ratio $m_{eff}/G\sigma$ was lower due to the small size of the string resonator and the use of high-stress stoichiometric SiN. Large portions of this study [22] were also conducted for driving amplitudes much smaller than $a_{crit}$, thus increasing the relative contribution of thermomechanical noise. In FIG. 5 (a), we also plot the actual predicted $NEP$ values by considering $\varepsilon_{\lambda,\theta}(\lambda, 0) \approx 0.4$ to illustrate the performance of the membrane at $\lambda \approx 12$ µm, which is the maximum sensitivity wavelength of a SiN membrane [18]. We note that $NEP_{th}$ (~$4.1 \times 10^{-11}$ W/$\sqrt{Hz}$) dominates especially at frequencies greater than $\pi f_r/Q$ (vertical yellow dashed line), i.e., where the thermomechanical noise is filtered out by the system mechanical response time.

Thermal fluctuation-dominated frequency noise, as demonstrated in FIG. 5 (a), is important as it could potentially enable thermal radiation sensors operating at the never achieved fundamental detectivity limit set by Eq. (12). In principle, doing so would only require functionalizing the membrane front side to become a perfect radiation absorber (i.e., $\varepsilon_{\lambda,\theta} = \varepsilon = 1$) and to thermally isolate the backside (i.e., $\varepsilon_{\lambda,\theta} = \varepsilon = 0$), while not deteriorating the membrane mechanical properties.

For practical situations, an important next step remains, i.e., predicting thermal fluctuation frequency noise after processing by the frequency tracking experimental apparatus (a digital PLL in our particular case). Noise described in our work Eqs. (10), (13) only include intrinsic resonator noise without the PLL contribution. As demonstrated in [20]–[22], PLL frequency tracking can amplify thermomechanical and detection noise by orders of magnitude, especially if the PLL bandwidth (50 Hz in the present work) is significantly greater than the resonator mechanical bandwidth (typically $1/2\pi\tau_{mech} \approx 0.03$ Hz in this example). This notably explains why our experimentally measured frequency noise ($6 \times 10^{-3}$ Hz RMS, see FIG. 2 b) is ~3 orders of magnitude greater than the frequency noise predicted by Eq. (10), (13), (~$10^{-6}$ Hz RMS). Considering the PLL loop dynamics along with detection and thermomechanical noise yields a noise prediction that is closer to reality (see supplementary information S1), even though we currently have no way to include thermal fluctuation noise in this analysis. This would obviously be a logical next step to our work.

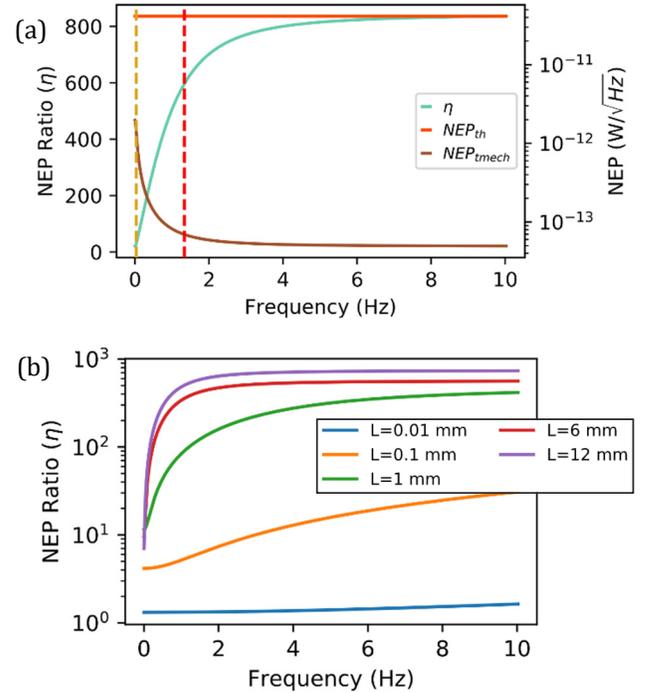

FIG. 5. Ratio of thermal NEP to thermomechanical NEP. (a) NEP for a typical membrane from this work ($d$ = 90 nm, $L$ = 6 mm, $Q \approx 10^6$, $\sigma$ = 100 MPa), which is theoretically dominated by thermal noise. Membrane mode (3,4) ($f_r \approx$ 76 kHz) and critical amplitude actuation are considered in this case. The red dashed line represents the bandwidth imposed by the thermal time constant of the membrane ($1/2\pi\tau_{th}$) while the yellow dashed line represents the resonator mechanical bandwidth ($1/2\pi\tau_{mech}$). (b) NEP ratio for additional membrane side-lengths ($L$ = 0.01 mm, 0.1 mm, 1 mm, 6 mm, 12 mm). Here we can see thermal fluctuation noise can dominate even with smaller non-radiation dominant membrane sizes.

## V. CONCLUSION

In the first portion of this work, after comparing two recently proposed models for the thermal transport in SiN resonators [8], [15], we find that they both correspond relatively well with experimental results when using the recently proposed values for thermal conductivity [17] of SiN, and the analytical emissivity model from [8], [18].

The circular approximation model [8] corresponds slightly better with experimental results, but differences are modest and are likely acceptable in any application.

Secondly, using our experimentally validated heat transport model, we show that thermal fluctuation noise can often dominate intrinsic frequency noise in SiN membrane resonators. This outlines the importance of considering this noise source in recently proposed formalism [20]–[22] for frequency noise in PLL frequency tracking sensing experiments, which so far focused on thermomechanical and detection noise. Ultimately, a complete understanding of thermal fluctuation frequency noise may lead to the demonstration of radiation sensors achieving the fundamental detectivity limit of thermal radiation sensors.

[1] H. Zhang and E. S. Kim, "Micromachined acoustic resonant mass sensor," *Journal of Microelectromechanical Systems*, vol. 14, no. 4, pp. 699–706, Aug. 2005, doi: 10.1109/JMEMS.2005.845405.

[2] M. S. Hanay *et al.*, "Single-protein nanomechanical mass spectrometry in real time," *Nature Nanotechnology*, vol. 7, no. 9, Art. no. 9, Sep. 2012, doi: 10.1038/nnano.2012.119.

[3] J. Chaste, A. Eichler, J. Moser, G. Ceballos, R. Rurali, and A. Bachtold, "A nanomechanical mass sensor with yoctogram resolution," *Nature Nanotechnology*, vol. 7, no. 5, Art. no. 5, May 2012, doi: 10.1038/nnano.2012.42.

[4] J. Moser *et al.*, "Ultrasensitive force detection with a nanotube mechanical resonator," *Nature Nanotechnology*, vol. 8, no. 7, Art. no. 7, Jul. 2013, doi: 10.1038/nnano.2013.97.

[5] D. Hälg *et al.*, "Membrane-Based Scanning Force Microscopy," *Phys. Rev. Applied*, vol. 15, no. 2, p. L021001, Feb. 2021, doi: 10.1103/PhysRevApplied.15.L021001.

[6] L. Laurent, J.-J. Yon, J.-S. Moulet, M. Roukes, and L. Duraffourg, "12 – μ m-Pitch Electromechanical Resonator for Thermal Sensing," *Phys. Rev. Applied*, vol. 9, no. 2, p. 024016, Feb. 2018, doi: 10.1103/PhysRevApplied.9.024016.

[7] M. Piller, N. Luhmann, M.-H. Chien, and S. Schmid, "Nanoelectromechanical infrared detector," in *Optical Sensing, Imaging, and Photon Counting: From X-Rays to THz 2019*, Sep. 2019, vol. 11088, p. 1108802. doi: 10.1117/12.2528416.

[8] C. Zhang, M. Giroux, T. A. Nour, and R. St-Gelais, "Radiative Heat Transfer in Freestanding Silicon Nitride Membranes," *Phys. Rev. Applied*, vol. 14, no. 2, p. 024072, Aug. 2020, doi: 10.1103/PhysRevApplied.14.024072.

[9] X. C. Zhang, E. B. Myers, J. E. Sader, and M. L. Roukes, "Nanomechanical Torsional Resonators for Frequency-Shift Infrared Thermal Sensing," *Nano Lett.*, vol. 13, no. 4, pp. 1528–1534, Apr. 2013, doi: 10.1021/nl304687p.

[10] N. Snell, C. Zhang, G. Mu, and R. St-Gelais, "Nanowatt Thermal Radiation Sensing using Silicon Nitride Nanomechanical Resonators," in *2020 Photonics North (PN)*, May 2020, pp. 1–1. doi: 10.1109/PN50013.2020.9167002.

[11] M. Giroux, C. Zhang, N. Snell, G. Mu, M. Stephan, and R. St-Gelais, "High Resolution Measurement of Near-Field Radiative Heat Transfer enabled by Nanomechanical Resonators," Jun. 2021, Accessed: Jul. 06, 2021. [Online]. Available: https://arxiv.org/abs/2106.09504v1

[12] L. Vicarelli, A. Tredicucci, and A. Pitanti, "Micromechanical bolometers for sub-Terahertz detection at room temperature," *arXiv:2107.12170 [physics]*, Jul. 2021, Accessed: Aug. 06, 2021. [Online]. Available: http://arxiv.org/abs/2107.12170

[13] B. M. Zwickl *et al.*, "High quality mechanical and optical properties of commercial silicon nitride membranes," *Appl. Phys. Lett.*, vol. 92, no. 10, p. 103125, Mar. 2008, doi: 10.1063/1.2884191.

[14] R. St-Gelais, S. Bernard, C. Reinhardt, and J. C. Sankey, "Swept-Frequency Drumhead Optomechanical Resonators," *ACS Photonics*, vol. 6, no. 2, pp. 525–530, Feb. 2019, doi: 10.1021/acsphotonics.8b01519.

[15] M. Piller *et al.*, "Thermal radiation dominated heat transfer in nanomechanical silicon nitride drum resonators," *Appl. Phys. Lett.*, vol. 117, no. 3, p. 034101, Jul. 2020, doi: 10.1063/5.0015166.

[16] A. Sikora *et al.*, "Highly sensitive thermal conductivity measurements of suspended membranes (SiN and diamond) using a 3ω-Völklein method," *Review of Scientific Instruments*, vol. 83, no. 5, p. 054902, May 2012, doi: 10.1063/1.4704086.

[17] H. Ftouni *et al.*, "Thermal conductivity of silicon nitride membranes is not sensitive to stress," *Phys. Rev. B*, vol. 92, no. 12, p. 125439, Sep. 2015, doi: 10.1103/PhysRevB.92.125439.

[18] C. Zhang, A. Bouchard, M. Giroux, T. A. Nour, and R. St-Gelais, "Erratum: Radiative Heat Transfer in Freestanding Silicon Nitride Membranes [Phys. Rev. Appl. 14, 024072 (2020)]," *Phys. Rev. Applied*, vol. 16, no. 1, p. 019901, Jul. 2021, doi: 10.1103/PhysRevApplied.16.019901.

[19] X. Zhang and C. P. Grigoropoulos, "Thermal conductivity and diffusivity of free-standing silicon nitride thin films," *Review of Scientific Instruments*, vol. 66, no. 2, pp. 1115–1120, Feb. 1995, doi: 10.1063/1.1145989.

[20] A. Demir, "Understanding fundamental trade-offs in nanomechanical resonant sensors," *Journal of Applied Physics*, vol. 129, no. 4, p. 044503, Jan. 2021, doi: 10.1063/5.0035254.

[21] A. Demir and M. S. Hanay, "Fundamental Sensitivity Limitations of Nanomechanical Resonant Sensors Due to Thermomechanical Noise," *IEEE Sensors Journal*, vol. 20, no. 4, pp. 1947–1961, Feb. 2020, doi: 10.1109/JSEN.2019.2948681.

[22] P. Sadeghi, A. Demir, L. G. Villanueva, H. Kähler, and S. Schmid, "Frequency fluctuations in nanomechanical silicon nitride string resonators," *Phys. Rev. B*, vol. 102, no. 21, p. 214106, Dec. 2020, doi: 10.1103/PhysRevB.102.214106.

[23] J. R. Vig and Yoonkee Kim, "Noise in microelectromechanical system resonators," *IEEE Transactions on Ultrasonics, Ferroelectrics, and Frequency Control*, vol. 46, no. 6, pp. 1558–1565, Nov. 1999, doi: 10.1109/58.808881.

[24] A. N. Cleland and M. L. Roukes, "Noise processes in nanomechanical resonators," *Journal of Applied Physics*, vol. 92, no. 5, pp. 2758–2769, Aug. 2002, doi: 10.1063/1.1499745.

[25] D. Rugar, H. J. Mamin, and P. Guethner, "Improved fiber-optic interferometer for atomic force microscopy," *Appl. Phys. Lett.*, vol. 55, no. 25, pp. 2588–2590, Dec. 1989, doi: 10.1063/1.101987.

[26] C. H. Mastrangelo, Y.-C. Tai, and R. S. Muller, "Thermophysical properties of low-residual stress, Silicon-rich, LPCVD silicon nitride films," *Sensors and Actuators A: Physical*, vol. 23, no. 1, pp. 856–860, Apr. 1990, doi: 10.1016/0924-4247(90)87046-L.

[27] G. Carlotti *et al.*, "Measurement of the elastic and viscoelastic properties of dielectric films used in microelectronics," *Thin Solid Films*, vol. 414, no. 1, pp. 99–104, Jul. 2002, doi: 10.1016/S0040-6090(02)00430-3.

[28] "Dielectric Constant & Relative Permittivity » Electronics Notes." https://www.electronics-notes.com/articles/basic_concepts/capacitance/dielectric-constant-relative-permittivity.php (accessed Oct. 19, 2020).

[29] M. Sansa *et al.*, "Frequency fluctuations in silicon nanoresonators," *Nature Nanotechnology*, vol. 11, no. 6, pp. 552–558, Jun. 2016, doi: 10.1038/nnano.2016.19.

[30] A. Rogalski, "Infrared detectors: status and trends," *Progress in Quantum Electronics*, vol. 27, no. 2, pp. 59–210, Jan. 2003, doi: 10.1016/S0079-6727(02)00024-1.

[31] S. Schmid, L. G. Villanueva, and M. L. Roukes, *Fundamentals of Nanomechanical Resonators*. Springer International Publishing, 2016. doi: 10.1007/978-3-319-28691-4.

# Heat Transport in Silicon Nitride Drum Resonators and its Influence on Thermal Fluctuation-induced Frequency Noise

# (Supplementary Information)

Nikaya Snell, Chang Zhang, Gengyang Mu, Raphael St-Gelais*
Department of Mechanical Engineering, University of Ottawa, Ottawa, ON, Canada
*raphael.stgelais@uottawa.ca

## S.1 PLL Noise Analysis

The measured experimental noise in the main text (main text Fig. 2 b), is higher than predicted by our equations for the intrinsic resonator noise (Eq. 11, 15) due to the use of a PLL frequency tracking scheme. Such behaviour is expected and predictable using the theory given in [20], [21], and is especially pronounced for PLL bandwidths that are higher than the resonator bandwidth $[BW = (\pi f_r)/Q]$. High PLL bandwidths (~50 Hz) are required in our case for accurately characterizing our membrane thermal response time (see main text Fig. 3). Incorporating PLL loop dynamics into our noise analysis is possible by computing the predicted Allan Deviation due to thermomechanical and detection noise as given in [20]–[22]:

$$\sigma_y(\tau) = \frac{2}{\sqrt{\pi}\tau} \left[ \int_{-\infty}^{+\infty} \frac{\left[\sin\left(\frac{\omega\tau}{2}\right)\right]^4}{\omega^2} S_y(\omega)\, d\omega \right]^{\frac{1}{2}}, \tag{S1}$$

where $\tau$ is the integration time and $S_y(\omega)$ is the spectral density of the fractional frequency fluctuations. Using Eq. (13) and (16) from the main text, the frequency noise from thermomechanical and detection noise is

$$S_y(\omega) = \frac{k_B T}{m_{eff}\omega_r^3 Q a_{rss}^2} |H_{tmech}^{PLL}(j\omega)|^2 + \kappa_d^2 \frac{k_B T}{m_{eff}\omega_r^3 Q a_{rss}^2} |H_{det}^{PLL}(j\omega)|^2, \tag{S2}$$

where $H_{tmech}^{PLL}$ and $H_{det}^{PLL}$ are transfer functions for thermomechanical noise and detection noise in a closed-loop system [20], [21]:

$$H_{tmech}^{PLL}(s) = \frac{(sK_p + K_i)H_L(s)}{s^2 + \frac{s}{\tau_{mech}} + (sK_p + K_i)H_L(s)}, \tag{S3}$$

$$H_{det}^{PLL}(s) = \frac{1}{H_{tmech}(s)} \frac{(sK_p + K_i)H_L(s)}{s^2 + \frac{s}{\tau_{mech}} + (sK_p + K_i)H_L(s)}. \tag{S4}$$

In the above expressions, two transfer functions are applied: the resonator filter $H_{tmech}(s) = \frac{1}{1+s\tau_{mech}}$ and the demodulator filter $H_L(s) = \frac{1}{1+s\tau_{demod}}$. Also introduced are the loop controller parameters $K_p$ (proportional) and $K_i$ (integral), which are calculated as a function of the PLL bandwidth [20], [21]:

$$K_p = \omega_{PLL}, \qquad K_i = \frac{\omega_{PLL}}{\tau_{mech}} \qquad (S5)$$

Note that Eq S1-S5 do not include thermal fluctuation-induced frequency noise $S_{y,th}(\omega)$ because a transfer function $H_{th}^{PLL}(s)$ for this noise term is unknown to the authors. As stated in the conclusion of our manuscript, deriving such term and studying thermal-frequency fluctuation in closed loop would be the logical next step to our work.

In FIG. S1 (a) we show a 5-minute frequency noise measurement (similar to Fig 2b of the main text, but for a longer acquisition time), which is then converted to an Allan deviation and plotted (FIG. S1 (b)) with the model detailed above. In FIG. S1 (b), we compare the experiment data with the Allan deviation for a 0.03 Hz PLL bandwidth, which would be optimally matched to the mechanical bandwidth of the resonator, and for a 50 Hz PLL bandwidth, which was the bandwidth used in our experiments to resolve the thermal response time. We can clearly see an increase in system noise by ~3 orders of magnitude for our 50 Hz PLL bandwidth, compared with the noise that would be expected if we had matched our PLL bandwidth to the resonator bandwidth. In the latter case, the Allan deviation closely follows the asymptotic behaviours for thermomechanical and detection noise (i.e., no extra noise is expected above the predictions of Eq. 13, 16). The noise present in our experiment more closely corresponds to the theoretical model for the 50 Hz PLL bandwidth. This supports the case that the higher noise levels seen experimentally are due to the use of a high PLL bandwidth. At $\tau \approx 0.03$ s, there is an increase in the experimental Allan deviation, which is not present in the model. The exact source of this increase is still unknown; however, a possible cause could be noise introduced while driving our membrane near the onset of nonlinearity. At higher integration times, noise due to drift dominates our membrane resulting in further deviation from the theoretical model, as is commonly seen experimentally [22].

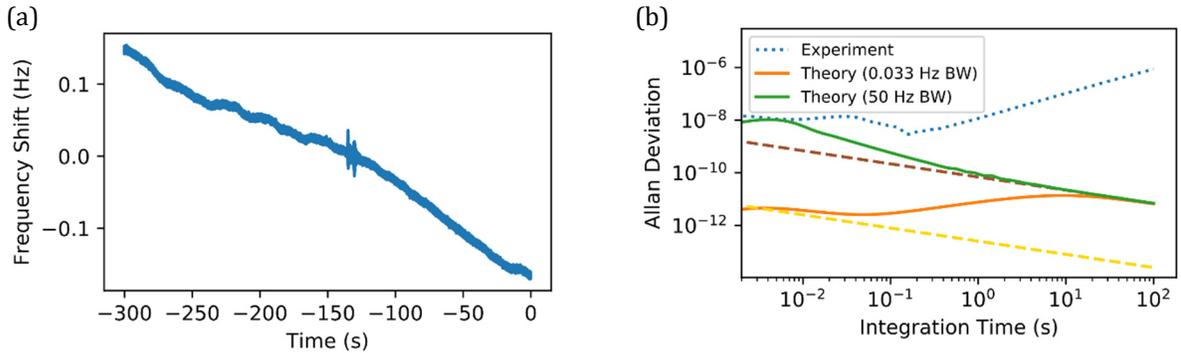

**FIG. S1**. (a) Frequency noise trace over a duration of 5-minutes for mode (3, 4) of an $L = 6$ mm membrane. (b) Theoretical Allan Deviations (solid lines) and experimental (dotted line) Allan deviations of mode (3,4) of an $L = 6$ mm membrane. Increasing the PLL bandwidth from 0.03 Hz to 50 Hz results in an increase in the Allan deviation by ~3 orders of magnitude. For a PLL bandwidth matched to the resonator bandwidth (0.03 Hz) the Allan deviation follows the detection noise asymptote (yellow dashed line) and the thermomechanical noise asymptote (brown dashed line)

To calculate the detection noise ($S_{y,det}$, Eq. 16 in main text) in our setup and generate the plot seen in FIG. S1 (b), we measured the dimensionless parameter $\kappa_d$, which indicates the relative contributions of thermomechanical and detection noise. This parameter was determined experimentally by measuring the power spectral density (PSD) over 100 Hz bandwidth near the resonance peak of the membrane when no drive was provided by the shear piezo in our setup. Under these conditions we can still see a peak in the PSD at the resonance frequency of our resonator due to thermomechanical noise. Taking the ratio between the $\sqrt{PSD}$ of the background detection noise (red dashed line in FIG. S2) and the $\sqrt{PSD}$ of the resonance peak induced by thermomechanical noise (green dashed line in FIG. S2) results in a $\kappa_d$ value of 0.006.

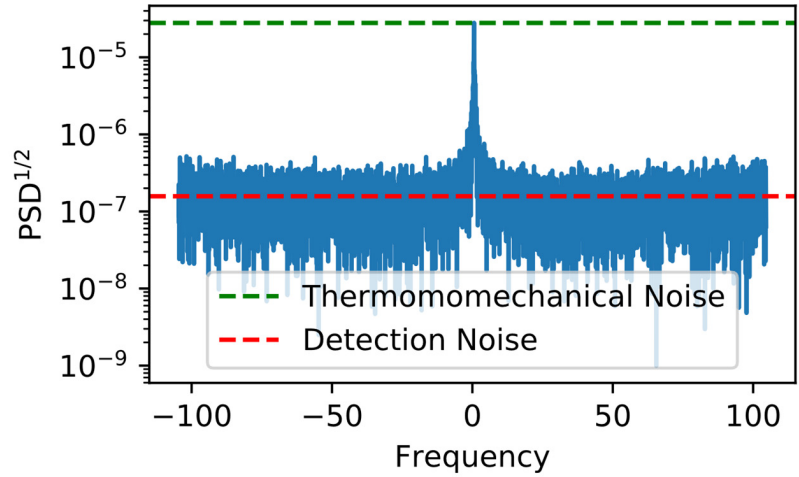

**FIG. S2**. Measurement of detection amplitude noise relative to thermomechanical noise ($\kappa_d$). Spectrum over 100 Hz BW near the resonance peak of the membrane is measured when no actuation is provided by the shear piezo (i.e., resonator motion is a result of random motion of particles around and within the membrane). The $\sqrt{PSD}$ ratio between the background detection noise (red dashed line) and the resonance peak caused by thermomechanical noise (green dashed line) yields $\kappa_d$ = 0.006.